\documentclass[twocolumn,9pt]{article}
\usepackage{extsizes}
\usepackage[super,sort&compress,comma]{natbib} 
\usepackage[left=1.5cm, right=1.5cm, top=1.785cm, bottom=2.0cm]{geometry}
\usepackage{balance}
\usepackage{mathptmx}
\usepackage{sectsty}
\usepackage{graphicx} 
\usepackage{lastpage}
\usepackage[format=plain,justification=justified,singlelinecheck=false,font={stretch=1.125,small,sf},labelfont=bf,labelsep=space]{caption}
\usepackage{float}
\usepackage{fancyhdr}
\usepackage{fnpos}
\newcommand{\ecoli}{{\it E. coli} }
\usepackage[english]{babel}
\addto{\captionsenglish}{%
  
}
\usepackage{array}
\usepackage{droidsans}
\usepackage{charter}
\usepackage[T1]{fontenc}
\usepackage[usenames,dvipsnames]{xcolor}
\usepackage{setspace}
\usepackage[compact]{titlesec}
\usepackage{xcolor}
\usepackage{hyperref}
\usepackage{subcaption}
\usepackage{epstopdf}

\newenvironment{changemargin}[2]{%
\begin{list}{}{%
\setlength{\topsep}{0pt}%
\setlength{\leftmargin}{#1}%
\setlength{\rightmargin}{#2}%
\setlength{\listparindent}{\parindent}%
\setlength{\itemindent}{\parindent}%
\setlength{\parsep}{\parskip}%
}
\item[]}{\end{list}}

\title{\LARGE{\textbf{Bacterial activity hinders particle sedimentation$^\dag$}}}
\author{Jaspreet Singh,Alison E. Patteson,Bryan O. Torres Maldonado,Prashant K. Purohit,Paulo E. Arratia}
\date{\today}

\begin{document}

\twocolumn[
 \begin{@twocolumnfalse}

\begin{center}
\noindent\LARGE{\textbf{Bacterial activity hinders particle sedimentation}}\\
\end{center}
\begin{changemargin}{+1.5cm}{+1.5cm}
  \noindent\large{Jaspreet Singh,\textit{$^{\ddag a}$}   Alison E. Patteson,\textit{$^{\ddag b}$} Bryan O. Torres Maldonado,\textit{$^{a}$} Prashant K. Purohit,\textit{$^{a}$}  and Paulo E. Arratia\textit{$^{a}$}} 
 \end{changemargin}
 \begin{changemargin}{+2.2cm}{+2.2cm}
 \textit{$^{a}$~Department of Mechanical Engineering \& Applied Mechanics,
		University of Pennsylvania, Philadelphia, PA 19104. E-mail:parratia@seas.upenn.edu}; \textit{$^{b}$ Dept. of Physics, Syracuse University, Syracuse, NY 13244}
	 \begin{center}	\ddag~Equal contribution
	\end{center} 

 \end{changemargin}
\begin{changemargin}{+1.5cm}{+1.5cm}
\noindent\normalsize{\\ Sedimentation in active fluids has come into focus due to the ubiquity of swimming micro-organisms in natural and industrial processes. Here, we investigate sedimentation dynamics of passive particles in a fluid as a function of bacteria {\it E. coli} concentration. Results show that the presence of swimming bacteria significantly reduces the speed of the sedimentation front even in the dilute regime, in which the sedimentation speed is expected to be independent of particle concentration. Furthermore, bacteria increase the dispersion of the passive particles, which determines the width of the sedimentation front. For short times, particle sedimentation speed has a linear dependence on bacterial concentration. Mean square displacement data shows, however, that bacterial activity decays over long experimental (sedimentation) times. An advection-diffusion equation coupled to bacteria population dynamics seems to capture concentration profiles relatively well. A single parameter, the ratio of single particle speed to the bacteria flow speed can be used to predict front sedimentation speed.} \\

\end{changemargin}
 \end{@twocolumnfalse} ]

\footnotetext{\textit{$^{a}$~Department of Mechanical Engineering \& Applied Mechanics, University of Pennsylvania, Philadelphia, PA 19104. E-mail: parratia@seas.upenn.edu}; \textit{$^{b}$ Dept. of Physics, Syracuse University, Syracuse, NY 13244}}
\footnotetext{\ddag~J.S.and A.P. contributed equally to this work.}

\section{Introduction}

The settling of organic and inorganic matter in fluids plays an important role in many technological and natural processes \cite{guazzelli2011physical,Hinch2010,davis1985sedimentation}. In industry, proper dispersion of particulates in liquids is essential to the production of foodstuff, paints, biofuels, and plastics.  In oceans, sedimentation of biological matter play an important role on the regulation of planktonic organisms' position relative to light and foraging strategies and is a key part of the ocean carbon cycle (i.e. ocean's biological pump) that transports carbon from the ocean's surface to depth \cite{treguer2000global,sarmiento1984model}. Recently, there has been much interest in the sedimentation of active particles, which are usually defined as self-propelling particles (living or synthetic) that inject energy, generate mechanical stresses, and create flows within the fluid medium \cite{marchetti2013}. These particles can drive the fluid out of equilibrium (even in the absence of external forcing) and lead to many interesting phenomena such as collective behavior \cite{marchetti2013,sriram2017}, unusual viscosity \cite{lopez2015,lindner2013}, and an enhancement in particle diffusivity \cite{Wu2000,Chen2007,Leptos2009,Mino2011,Jepson2013} that depends anomalously on particle size \cite{Patteson2016,brady1988sedimentation}. Describing such active systems remains challenging, particularly under the effects of external forcing such as gravity \cite{Cates2008,Cates2009,Cates2010,PattesonOpinion}.
 
Recent studies have mainly focused on the \textit{steady-state} sedimentation of suspensions of active particles. Experiments with dilute active colloids such as phoretic particles found that density profiles at steady state decay exponentially with height yielding a sedimentation length that is larger than that expected for thermal equilibrium \cite{Palacci2010,Ginot2015}; similar results are found even when the sedimentation speed is of the same order as the particle propulsion speed \cite{ginot2018sedimentation}. This enhancement can be described by an effective activity-induced temperature that correlates with the particle's ability to self-propel and achieve larger diffusivities than from thermal fluctuations alone. These results agree relatively well with theory \cite{Cates2008,Cates2009} and simulations \cite{Cates2010, Tsao2014} for active particles that are either non-interacting \cite{Cates2008,Cates2009} or with limited hydrodynamic interactions \cite{Cates2010, Tsao2014}. 
\begin{figure*}
 \centering
 \includegraphics[height=5cm]{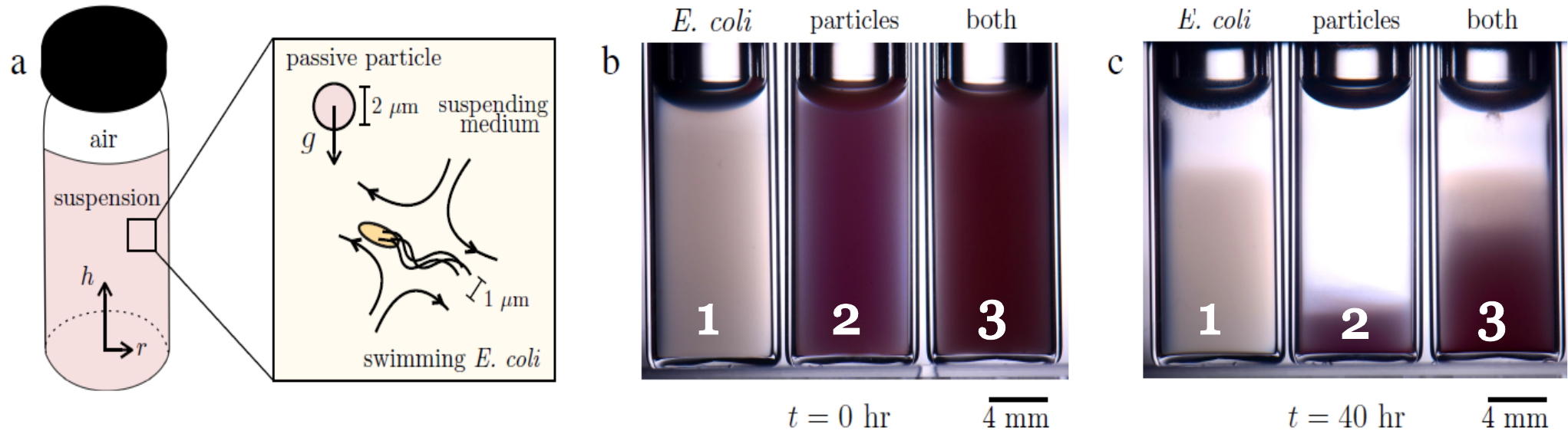}
 \caption{Experimental setup and sample images: (a) A schematic of the setup and bacteria/particle suspensions. Sedimentation experiments are conducted in sealed glass vials that include a volume of atmospheric air. The particles are 2 $\mu$m polystyrene spheres, subject to gravity. The bacteria are 2 $\mu$m rod-shaped \emph{E. coli}, which generate local extensile fluid flows when swimming. Samples are uniformly mixed at the start of the experiments. (b) A sample experiment shows three representative samples: suspensions of (i) only \emph{E. coli} ($\phi_b=0.24 \%$), (ii) only particles ($\phi_p=0.04 \%$), and (iii) \emph{E. coli} and particles ($\phi_b=0.24 \%+\phi_p=0.04 \%$) at $t=0$ hr, the start of the experiment. (c) After 40 hours, the samples have sedimented to various heights. The passive particles sediment much faster than the \emph{E. coli}. When particles and $E.$ $coli$ are combined, the passive particles (pink) extend to higher heights than in the absence of bacteria.}
 \label{Fig_1}
\end{figure*}

Experiments with swimming micro-organisms, however, paint a more nuanced picture. Under an external centrifugal field, \textit{Escherichia coli} (\textit{E}. \textit{coli}) fractionizes by motility so that fast-swimming bacteria swim throughout the sample and slow-swimming bacteria accumulate at the bottom; the resultant particle distribution matches a model of active colloids that possess a spectrum of effective temperatures \cite{Leonardo2013}. In the presence of extra-cellular polymers, it has been found that bacteria can aggregate and thus enhance sedimentation rates \cite{Poon2012}; however motile bacteria are more resistant to this aggregation than non-motile bacteria due to their enhanced diffusivity. In mixtures of swimming algae and passive particles, the steady-state sedimentation profile of passive particle is found to be described by an effective diffusivity (or temperature) that increases linearly with the concentration of swimming microbes \cite{Polin2016}. While the concept of effective temperatures and enhanced diffusivities have been useful in describing the steady-state sedimentation profiles of active systems, the transient unsteady evolution of such active systems remains largely unknown.  How a distribution of an initially homogeneous mixture of active and passive particle suspension subject to gravity change over time is a question that remains unanswered.

In this manuscript, we investigate the sedimentation dynamics of bacterial suspensions in experiments and in a simple model. Active suspensions are prepared by mixing {\it E. coli}, a model biological organism widely used for motility \cite{berg2008}, and polystyrene colloidal particles in buffer solutions. We study these initially well-mixed suspensions as they settle over relatively long periods of time (up to 72 hours) and use image analysis techniques to track the evolving density profile and the spreading interface at the top of the settling particle suspension (Fig. \ref{Fig_1}). Our results show that the presence of bacteria hinders (passive) particle sedimentation speed and increases their macroscopic dispersion. On the other hand, bacteria sedimentation speed remains unaffected by the presence of passive particles in concentration range investigated here. At long times, the particle concentration profiles can be significantly affected by the appearance of dead bacteria due to finite levels of nutrients and oxygen in our bottles. These effects can be captured using an advection-diffusion equation coupled with bacteria population dynamics. Moreover, the sedimentation process can be captured relatively well by the ratio of two main speeds, namely the particle suspension mean sedimentation speed and the bacterial flow speed (\emph{cf.} Fig. \ref{scaling_fig}). 
\section{Experimental Methods}
The experimental fluids are suspensions of swimming  {\it Escherichia coli} (wild-type K12 MG1655) and passive polystyrene particles in a buffer solution (67 mM of NaCl in water). The bacterium {\it E. coli} is a model organism for flagellated bacteria motility and achieves net propulsion by rotating its helical flagella at approximately 100 Hz, driving the cell body forward at speeds of $10$-$20$ $\mu$m/s \cite{berg2008}. The (time-averaged) flow generated by swimming \emph{E. coli} are well approximated by a force dipole that decays with the distance from cell body $r$ as 1/$r^2$\cite{Goldstein2011}. Here, bacteria are grown to saturation ($10^9$ cells/mL) in culture media (LB broth, Sigma-Aldrich). The saturated culture is gently cleaned by centrifugation and is suspended in buffer at concentration $c$ ranging from 0.75{$\times 10^9$} to 7.5 $\times 10^9$ cells/mL. These concentrations are considered dilute, corresponding to volume fractions $\phi_{b} = c v_{\mathrm{b}}$ ranging from 0.1\% to 1\%, where $v_{\mathrm{b}}=$ 1.4 $\mu$m$^3$ is the  \emph{E. coli} body's volume~\cite{Jepson2013}. We do not observe any large scale collective behavior in these particle/bacteria suspensions, which is consistent with previous predictions and measurements on the concentration of bacteria ($\approx 10^{10}$ cells/mL) for the onset of collective motion \cite{Kasyap2014}.   Polystyrene spheres (Sigma Aldrich) with a diameter $d$ of $2$ $\mu$m and $\rho$ of 1.05 g/cm$^3$ are used as passive particles. Polystyrene particles are cleaned by centrifugation and then resuspended in the buffer-bacterial suspension. Particle concentrations are dilute at $1.0 \times 10^8$ particles/mL, which corresponds to $0.04 \%$ volume fraction and is kept fixed for all experiments shown here.

Sedimentation experiments are performed by introducing 1.5 mL of the fluid suspensions into glass vials  (8.3 mm in diameter, 20 mm in height), as shown schematically in Fig.~\ref{Fig_1}(a). The suspensions are gently mixed by hand with a pipette so that the particles are uniformly distributed at the start of the experiment ($t=0$ hr). The vials are capped and air volume (approximately {175} mm$^3$) remains inside of them. In order to reduce the light diffraction from the round vials and to control temperature, the samples are placed in a cube-shaped water bath maintained at $T_{0}=295$ K; round vials are used to avoid effects from sharp edges. Images are taken every 1 to 10 minutes for up to 7 days {with a Nikon D7100 camera that is equipped with a 100 mm Tokina lens}. The light source is a camera flash kit (Altura Photo) positioned behind the sample. 

We characterize the sedimentation processes by measuring the (i) the sedimentation (downward) speed $v$ of the passive particle supernatant-suspension interface and (ii) evolution of the particle concentration as a function of time $t$ and distance along the height of the vial $h$ (Fig.~\ref{Fig_1}a). Front sedimentation speeds are obtained using methods detailed in \cite{guazzelli2011physical}. The sedimentation speed of a single polystyrene particle in a viscous fluid of viscosity $\mu$ is estimated by considering a force balance of gravity and viscous drag acting on the particle. This yields  $v_s = (\Delta \rho) g d^2/{18 \mu}$, where $(\Delta \rho)$ is the density difference between the particle (1.05 g/cm$^3$) and suspending liquid (1.00 g/cm$^3$), $g$ is the acceleration due to gravity ($g = 9.81$ m/s$^2$), and $d$ is the particle diameter. For the 2 $\mu$m polystyrene particles in water, the sedimentation speed $v_s$ is $0.13~\mu$m/s.

To estimate particle concentrations along the height $h$ of the bottle, we use image analysis methods to obtain the variations in the intensity of the light transmitted $I (h)$ through the specimen -- the intensity of the transmitted light $I (h)$ is inversely proportional to the  concentration of passive particles and bacteria at that $h$. We select image intensity profiles as a function of height from the middle of the vial, far from the boundaries of the wall to avoid image aberrations. The image intensity profiles are then converted to particle number density through an intensity-density calibration curve, which is determined by measuring the image intensity of suspensions at known concentrations of passive particles and swimming bacteria. The resultant number densities are then multiplied by the volume of the individual particle to obtain the volume fraction as a function of height $h$ (\emph{cf.} Figs. 2).  

\section{Results and Discussion}

The main goal of this manuscript is to investigate the effects of biological activity on the sedimentation of passive particles. Figures \ref{Fig_1}(b) and (c) show snapshots of fluid suspensions taken at $t = 0$ hr (start of the experiment) and $t=40$ hr, respectively. The samples in Figure \ref{Fig_1}(b) and (c) correspond to, from left to right: (1) a suspension of only $E.$ $coli$ ($\phi_{b0} = 0.24\%$), (2) a suspension of only passive particles ($\phi_{p0}=0.04\%$), and (3) a suspension of passive particles and $E.$ $coli$ ($\phi_{p0}=0.04\%$, and $\phi_{p0}=0.24\%$ respectively). All samples exhibit a sedimentation front -- an interface between the aqueous supernatant at the top and the particulate suspension at the bottom -- that moves downwards from the top of the container at a certain sedimentation speed. The snapshots in Figs. 1(b) and (c) show that the \textit{E. coli} suspension (bottle 1) settles at a much lower rate than (passive) particle suspension (bottle 2), which demonstrates that activity can have a strong effect on sedimentation. 

Indeed, the sedimentation of passive particles in the presence of swimming bacteria (bottle 3) is significantly different from the sedimentation of the passive particles alone (bottle 2). We find that the sedimentation of the passive particles is hindered once bacteria are introduced to the passive suspension. The snapshots show that (i) passive particles (pink) are suspended for longer times (at higher heights) in the presence of bacteria and that (ii) the sedimentation front seems more dispersed compared to the sharp front observed in the absence of active bacteria (bottle 2). Overall, these results show that while the addition of bacteria can significantly affect the passive particle sedimentation process (Fig. 1b -- see passive particle front position in bottle 3 versus bottle 2), passive particles do not seem to affect bacteria sedimentation (Fig. 1b -- see \textit{E. coli} front position in bottle 3 versus bottle 1). In what follows, we will investigate these observations in more detail by systematically changing the bacteria concentration while maintaining the passive particle concentration constant at $\phi_{p0} = 0.04 \%$ for all experiments.  That is, we will systematically perturb the passive particle suspension with different levels of (bacterial) activity. In our experiments, the ratio of bacteria to passive particle initial concentration, $\xi=\frac{\phi_{b0}}{\phi_{p0}}$, ranges from 0.28 to 22.9; we note that all solutions are still considered to be in the dilute regime.
\begin{figure}
 \centering
 \includegraphics[height=7.5cm]{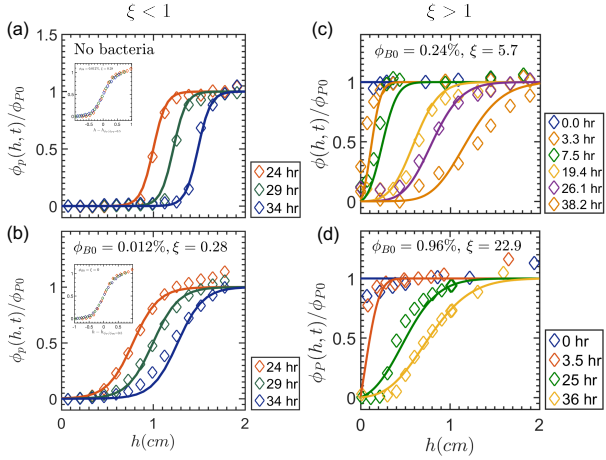}
 \caption{(a,b) Concentration profiles for low concentration ratios ($\xi<1$). Diamonds are the experimental data and the solid lines are from the solution to Eqn.\ref{conv_diff}. Suspensions with  (a) no or (b) low bacteria ($\phi_{b0}=0.012\%$) concentration can be adequately described using the Burgers' equation (see SM). The fitted dispersivities are$\mathcal{D}=0.75\mu m^2/s$ for (a) and $\mathcal{D}=1.50\mu m^2/s$ for (b). The presence of live bacteria increases the dispersivity by a factor of 2, while the front propagation speed $v\approx 0.12$ $\mu$m/s remains relatively constant. Insets show re-scaled profiles. (c,d) Sedimentation profiles for high concentration ratios ($\xi>1$); (c) $\phi_{b0}=0.24\%$ and (d) $\phi_{b0}=0.96\%$. Note that the dispersivities $\mathcal{D}_p=40 \mu m^2/s$ and $\mathcal{D}_p=80 \mu m^2/s$, respectively increase with the concentration of bacteria and are dramatically larger than the ones obtained in (b).  We obtain these profiles by integrating Eqns. (\ref{conv_diff_bact}) and (\ref{conv_diff_pass}).}
 \label{Fig_prof_low}
\end{figure}
\subsection{Low concentration ratio (\texorpdfstring{$\xi<1$}{Lg})} 
We now consider experiments where the bacteria to particle concentration ratio, $\frac{\phi_{b0}}{\phi_{p0}} = \xi$, is less than 1. To characterize the sedimentation process, we measure the particle concentration $\phi_p(h,t)$ as a function of distance along the bottle height $h$ and time $t$. Figure \ref{Fig_prof_low}(a) shows the normalized concentration profiles, $\phi_p(h,t)/\phi_{p0}$, for the passive particle case (no bacteria) as a function of bottle height $h$, where $\phi_{p0}~(=0.04\%)$ is the particle initial concentration. The $\phi_p(h,t)/\phi_{p0}$ profiles are plotted for three different times, $t=24$ hours, $t=29$ hours, and $t=34$ hours. We find that the profiles are characterized by distinct sigmoidal shapes, which translate in a roughly similar manner as the sedimentation process evolves \cite{martin}; the initial rise in concentration profiles, where the concentration changes abruptly, indicates the position of the sedimentation front. The measured shape of the concentration profiles for this case is consistent with previously measured profiles in passive suspensions of thermal \cite{Piazza2008} and athermal spherical particles \cite{Davies1988,Lee1992, Salin1994,brzinski2018observation}. The width of the sedimentation front is related to particle dispersivity, which for small particles in suspension is in part due to thermal motions and in part due to dispersion from long-range hydrodynamic interactions between multiple particles \cite{Ham1988,Chaikin1992,Davies1996,Nicolai1995}.

\begin{table}[h]
\small
  \caption{Some symbols used in sec.A}
  \label{table: symbols_small_conc}
  \begin{tabular*}{0.48\textwidth}{@{\extracolsep{\fill}}lll}
    \hline
   Symbol & Description  \\
    \hline
    $h$ & Coordinate along the height of the vial \\
     $\phi_{b}$ & Concentration of \ecoli  \\
    $\phi_p$ & Concentration of passive particles\\
    $\phi_{b0}$ & Initial concentration of live bacteria at $t=0$\\
    $\phi_{p0}$ & Initial concentration of passive particles at $t=0$\\
	$\mathcal{D}$ & Dispersivity of the passive particle front\\
	$v$ & Front propagation speed \\
	$\xi$ & $\frac{\phi_{b0}}{\phi_{p0}}$\\
    \hline
  \end{tabular*}
\end{table}

We can describe the concentration profiles of the passive particles in our control case--- passive particle suspension with no bacteria--- shown in Fig. 2(a) using an advection-diffusion equation of the form

\begin{equation}
	\frac{\partial \phi_p}{\partial t} + \frac{\partial (\phi_p~v(\phi_p))}{\partial h} = \frac{\partial}{\partial h}\left(\mathcal{D}\frac{\partial \phi_p}{\partial h}\right).
	\label{conv_diff}
\end{equation}

Here, $v$ is the speed of the sedimentation front and $\mathcal{D}$ is the particle suspension dispersivity. Due to hydrodynamic interactions between the settling particles, the speed of the sedimentation front $(v)$ is less than the terminal velocity of a single polystyrene particle ($v_s$ $\approx 0.13~\mu$m /s). This phenomenology is often described by a dimensionless hindering settling function, $H(\phi)=v(\phi_p)/v_s < 1$ \cite{richardson1954sedimentation,brzinski2018observation}. While there is still much debate on the exact form of $H(\phi)$, it has been recently shown that the Richardson-Zaki (RZ) formulation  $H(\phi)=v(\phi_p)/v_s=(1-\phi)^n$ \cite{richardson1954sedimentation} is able to describe the sedimentation of both Brownian ($n\approx 5.5$) and non-Brownian particles ($n\approx 4.5$) for a wide range of particle concentrations \cite{brzinski2018observation}; for very dilute suspension, $\phi < 0.04$ , both branches can be described by Batchelor's formulation with n=6.5 relatively well \cite{batchelor_1972}. Since the highest volume fraction (particles plus bacteria) is 1.04\% (or 0.0104), we adopt n=6.5. For such dilute suspensions, one can linearize the RZ expression such that $H(\phi)=v(\phi_p)/v_s \approx (1-n\phi_p)$, which is reminiscent of Batchelor's formulation \cite{batchelor_1972}.  This linearization permits us to transform the advection-diffusion equation (Eq.\ref{conv_diff}) into the well-known Burgers' equation which can be solved analytically \cite{martin} to obtain $v(\phi_p)$ (see SM). For Eq. \ref{conv_diff}, the initial condition is $\phi_{p}(h,t=0) = \phi_{p0}=0.04~\%$ and the only fitting parameter is the particle dispersivity $\mathcal{D}$.  

The solid lines in Fig. 2(a) show the best fit of Eq. \ref{conv_diff} to the passive particle sedimentation data, with $\mathcal{D}=0.75$ $\mu m^2/s$.  Relatively good agreement is found between the experimental data (diamonds) and the analytical results (solid lines) even at $t$=34 hours. The fitted dispersivity $\mathcal{D}=0.75$ $\mu m^2/s$ for the particle suspension is greater than the thermal diffusivity for a single sphere given by the Stokes-Einstein relation $D_0=k_B T/3 \pi \mu d=0.2~\mu m^2/s$ \cite{Einstein1905} , where $d=2$ $\mu$m is the diameter of the sphere, $k_B$ is the Boltzmann constant, $\mu$ is the fluid viscosity, and $T$ is the temperature ($T=295$ K). As mentioned before, the fitted dispersivity has contributions from thermal motions as well as from the long-range hydrodynamic interactions.  We note that although we used a linear function $v(\phi_p)/v_s = 1-n\phi_p$ to describe the hindered settling, our analysis indicates that the correction  $n\phi_p\ll 0.1$ is quite small, and the solution of Eq.\ref{conv_diff} for $\phi_{p0}=0.04~\%$ is almost identical when $v(\phi_p) \approx v_s =0.13 ~\mu$m/s. 

Next, we perturb the passive particle ($\phi_p=0.04\%$) case by adding a small amount (volume fraction $\phi_{b0}=0.012\%$) of live \ecoli into the vial.  Figure \ref{Fig_prof_low}(b) shows the experimentally measured (diamonds) normalized concentration profiles $\phi_p(h,t)/\phi_{0}$, for the active suspension as a function of height $h$; concentration profiles are measured for three different times, $t=24$ hr, $t=29$ hr, and $t=34$ hr. Similarly to the control case (i.e. passive particles), we find that the shape of the concentration profiles is characterized by a distinct sigmoidal jump. However, concentration jumps or transitions are less sharp indicating a broadening of the sedimentation fronts; active suspension normalized concentration profiles show smaller slopes than the passive particles case. 

Since the concentrations of both bacteria and passive particles are quite dilute and the shape of the profiles resemble the control case, we attempt to describe the sedimentation process using Eq.\ref{conv_diff}. However, we now have two fitting parameters, namely  $\mathcal{D}$ and $v$ (since we cannot assume a value for $n$ as before). The solid lines in Fig, 2(b) shows the best fit of Eq.\ref{conv_diff} to the experimental data with $v = 0.12~\mu m /s$ and $\mathcal{D} = 1.5~\mu m^2/s$. Overall Eq.\ref{conv_diff} is able to capture the normalized concentration profiles relatively well, but we do observe small deviations at long times ($t=34$ hr). Interestingly, while the front sedimentation speed remains nearly identical to the passive (control) case $v(\phi_p) \approx v_s$ = 0.13$\mu$m/s, the front dispersivity $\mathcal{D}$ increases two-fold from $0.75~\mu m^2$/s (passive) to $1.5~\mu m^2$/s (active). This is likely due to the bacterial swimming motion which can act to randomize and further spread particles in  the sedimentation front. Nevertheless, our results show that the macroscopic features of the sedimentation process of (very) dilute active suspensions, such as front sedimentation speed and dispersion coefficients, can still be described relatively well by an advection-diffusion equation with a constant $v$ and $\mathcal{D}$, particularly when the concentration of live bacteria is small or comparable to the concentration of passive particles i.e. $\xi < 1$. Next, we explore how the sedimentation of passive particles is affected as bacteria concentration is further increased (dilute nevertheless) and whether or not the above analysis remains adequate. 

\subsection{High concentration ratio (\texorpdfstring{$\xi>1$}{Lg})} 
We now investigate the cases in which bacteria (\ecoli) concentration is larger than the particle concentration such that $\xi=\phi_{b0}/\phi_{p0} >1$. We note that the system is overall still dilute and no collective motion is observed. As the concentration of the live \ecoli increases, we observe deviations from the  suspension without bacteria case, as shown in Fig. \ref{Fig_prof_low} (c) and (d) for $\xi$ = 5.7 and $\xi$= 22.9 respectively. Our experimental data (symbols in Fig. 2c,d) shows that, while the concentration profiles still show sigmoidal forms, the evolution of the concentration profiles does not quite follow the  self-similar behavior (Fig. 2a,b - inset) characteristic of the $\xi <1$ cases. Not surprisingly, Eq. \ref{conv_diff} fails to adequately describe the behavior of the suspension; not shown. What could be the causes for the observed deviations in the sedimentation dynamics?\\ 
\begin{figure*}[t!]
	\centering
	\begin{subfigure}[t]{.4\textwidth}
		\centering
		\includegraphics[width=1.0\textwidth, height=6cm]{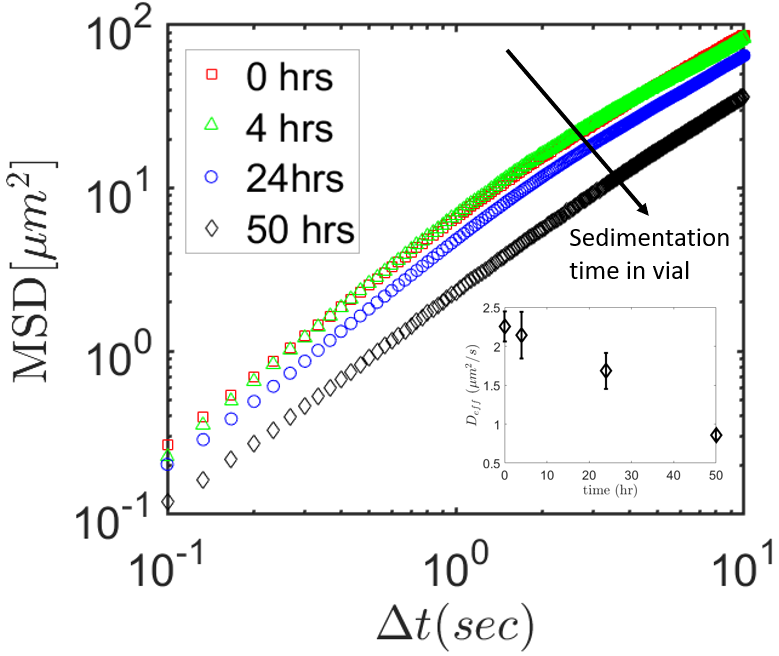}
		\caption{}
	\end{subfigure}
	\begin{subfigure}[t]{.4\textwidth}
		\centering
		\includegraphics[width=0.97\textwidth, height=6 cm]{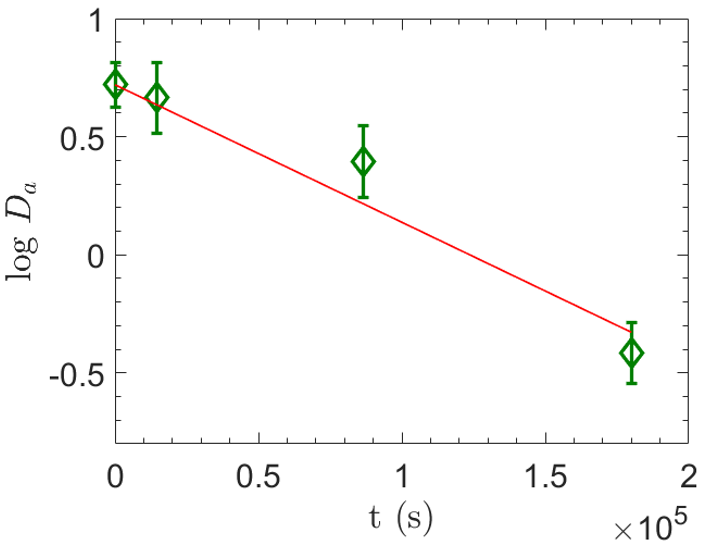}
		\caption{}
	\end{subfigure}
	\caption{(a) Mean square displacement (MSD) for an active suspension ($\phi_{b0}=0.31\%$ or $\xi = 7.4$) as a function of sedimentation time at $t=0,4,24$ and $50$ hr. MSD decreases with sedimentation time indicating suspension loss of activity or motility. Inset: Effective diffusivity $D_{eff}$ (see SM) as a function of time, showing the decrease in activity. (b) Active diffusivity, $D_a$, as a function of sedimentation time. Similar decrease is found using the expression $D_{eff}=D_0+D_a$, where $D_0$ is the fluid bare diffusivity (see ref.\cite{Patteson2016}). We use the decay in $D_a$ to obtain the bacteria loss of motility rate $k=5.7\times 10^{-6}/s$ (see text).} 
	\label{Fig_msd_D_t}
\end{figure*}

\begin{table}[h]
\small
  \caption{Concentrations of live bacteria used in the experiments. We distinguish the two regimes--low and high concentrations of live bacteria, by a parameter $\xi=\frac{\phi_{b0}}{\phi_{p0}}$. We show that when $\xi<1$ Burger's equation with increased dispersivity describes the concentration profiles. When $\xi>1$, the population dynamics of the bacteria needs to be accounted for}
  \label{table 2}
  \begin{tabular*}{0.48\textwidth}{@{\extracolsep{\fill}}lll}
    \hline
 		Cells per mL & Volume fraction $\phi_b$ ($\%$) &  $\xi=\frac{\phi_{b0}}{\phi_{p0}}$\\
    \hline
 
 			$0.75\times 10^8$  & $0.012$ & $0.28$\\
		 $1.5\times 10^9$ & $0.24$   & $5.7$\\
		 $3.0\times 10^9$ & $0.47$   & $11.4$\\
		 $4.5\times 10^9$ & $0.71$   & $17.1$ \\
		 $6.0\times 10^9$ & $0.94$   & $22.9$ \\
    \hline
  \end{tabular*}
\end{table}

We hypothesize that the deviations from the control case are due to \ecoli bacteria loss of activity or motility over time in the sealed vial due to nutrient depletion; experiments with bacteria can be quite long (up to 72 hours), and bacteria may run out of nutrients and oxygen. To test this hypothesis, we measure the mean square displacement (MSD) of passive particles (2 $\mu$m in diameter) in the presence of swimming bacteria to compute their effective diffusivity $D_{eff}$ as a function of sedimentation time in the vial. Here, we define the mean-squared particle displacement as MSD$(\Delta t)$ = $\langle|\textbf{r}(t_R + \Delta t) - \textbf{r}(t_R)|^2\rangle$, where the brackets denote an ensemble average over particles and reference times $t_R$. In short, we prepare several copies of the active suspensions (passive particles plus bacteria) and introduce them into several vials. We then withdraw $2~\mu L$ of fluid from a single vial at time $t=t_i$; the vial is then discarded. The withdrawn fluid is then stretched into a thin film using an adjustable wire frame with a thickness of $100$ $\mu m$; more information about this methodology can be found in \cite{Patteson2016}. We then track the passive particle displacement, $\textbf{r}$, to compute the MSD as a function of elapsed time $\Delta{t} \approx 10$ s.  We fit the MSD data to a generalized Langevin equation to obtain values of $D_{eff}$ as a function of (sedimentation) time $t$; see SM for more information.

Figure \ref{Fig_msd_D_t}(a) show the passive particle MSD data as a function of time for the $\xi = 11.4$ case at $t=0,4,24$ and $50$ hr in the vial. All curves show that, for long $\Delta t$, the MSD is linearly related to the time $\Delta t$. Importantly, we observe that the MSD curves decrease systematically with time, which indicates that bacteria are losing activity during the sedimentation process. This can be further quantified by plotting the measured $D_{eff}$ , which shows significant decrease with time as shown  in Fig. 3a (inset). 
To gain further insights into the effects of sedimentation time on bacteria activity, we compute the active component of the diffusion coefficient. In the dilute regime, $D_{eff}$ can be expressed as the sum of the Stokes-Einstein or bare diffusivity $D_0$ and an active diffusivity $D_\textrm{a}$. In dilute suspensions, $D_\textrm{a}$ increases linearly with the bacterial concentration \cite{Jepson2013,Kasyap2014,Patteson2016} such that

\begin{equation}
	D_{eff}(t)=D_0+D_1 \phi_{b,l}(t),
	\label{eq:diffu_eqn}
\end{equation}
where $D_1$ is a concentration-dependent active diffusivity and $\phi_{b,l}(t)$ is the concentration of the live \ecoli in the vial at time $t$. Since $D_0$ can be calculated and $D_{eff}$ can be measured, one can compute the values of the active diffusivity $D_a$ or $D_1$. Figure \ref{Fig_msd_D_t}b shows values of $D_a$ as a function of time for an active suspension at $\xi = 11.4$, and we observe a nearly exponential decay. In summary, the MSD data indicates that bacteria activity is diminish during the sedimentation process, i.e. bacteria are becoming non-motile and possibly dying. 

\begin{figure*}[]
	\centering
\centering
\begin{subfigure}[t]{.4\textwidth}
	\centering
	\includegraphics[width=0.8\textwidth, height=0.7\textwidth]{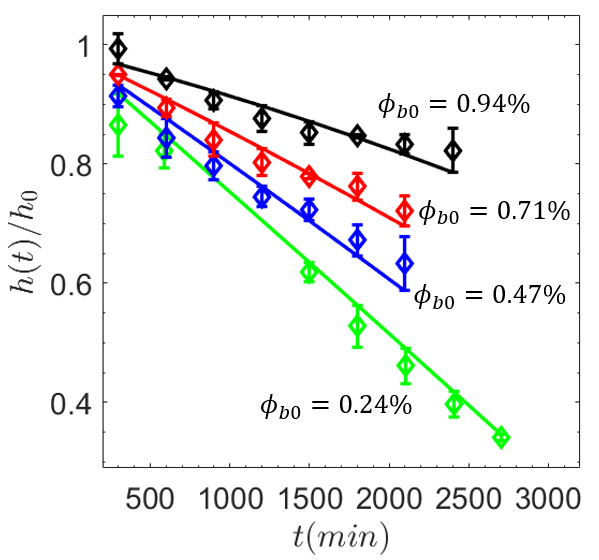}
	\caption{}
\end{subfigure}
\begin{subfigure}[t]{.4\textwidth}
	\centering
	\includegraphics[width=0.8\textwidth, height=0.7\textwidth]{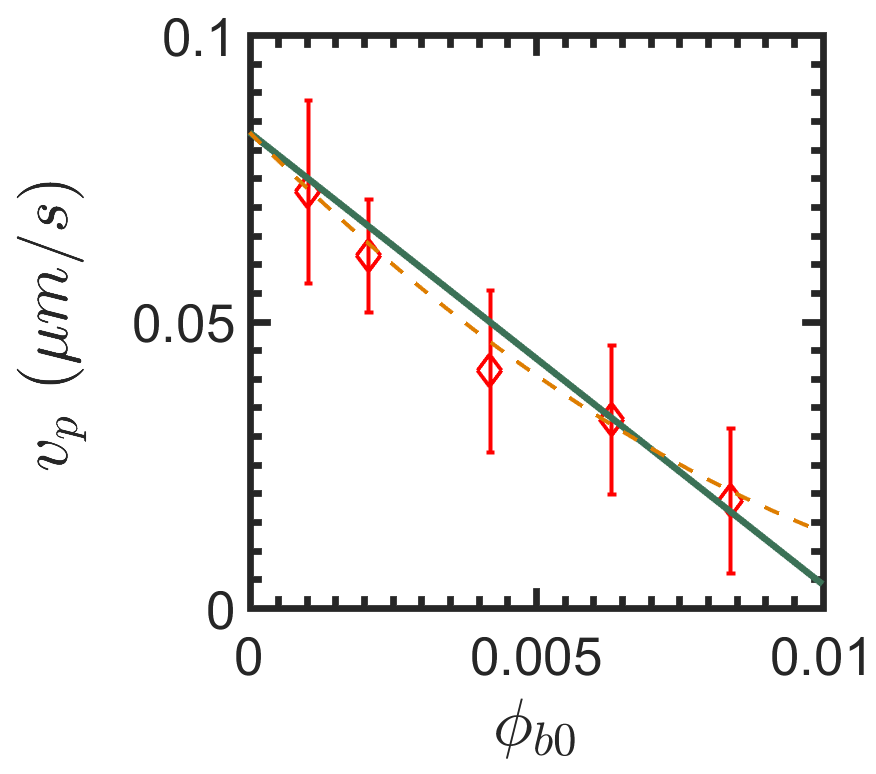}
	\caption{}
\end{subfigure}	\caption{(a) Position of particle sedimentation front, $h(t)/h_0$, where $h_0$ is the front initial concentration, for active suspensions (particles plus bacteria) as a function of time for a range of initial bacterial concentrations, $\phi_{b0}$. Data (open symbols) shows the decrease in front speed with increase in $\phi_{b0}$, as well as the initial linear dependence of sedimentation front with time. Solid lines are obtained by integrating the front position  $h(t)=h_0-v_s(t-n\phi_{b0}\frac{1-e^{-kt}}{k})$using $n$ and $v_s$ values obtained in (b), which yields an estimate of $k=1.0 \times 10^{-6}$/s. (b) Particle sedimentation speed $v_p$ as a function of bacterial concentration, $\phi_{b0}$.  The linear dependence is reminiscent of Batchelor's expression of the form $v_p(\phi_{b0})=v_s(1-n\phi_{b0})$ (\cite{batchelor_1972}. (Note that since the data is collected at initial times when $kt\ll1$ and $e^{-kt}\approx 1$, we can assume that $\phi_{b,l}\approx \phi_{b0}$.) Solid line shows best linear fit with $n\approx 120$ and $v_s \approx 0.1 \mu m/s$. A quadratic expression of the form$v_p (\phi_{b0})=v_s[1-n\phi_{b0} + ((n/2)\phi_{b0})^2]$ (dotted line), where $n \approx 120$ yields a slightly better fit alluding to the presence of second order effects. However, for the rest of the paper, we use the linear expression.} 
	\label{fig_vel}
\end{figure*}
Since an exponential decay is observed, the change in bacteria activity can be described via a first-order process,

\begin{equation}
\frac{d \phi_{b,l}}{d t}=-k\phi_{b,l}.
\label{eq:death_bac}
\end{equation}
where the constant $k$ can be thought of as bacteria loss of motility (or activity) rate. Here, we assume that the concentration of live \textit{E. coli}, $\phi_{b,l}$, is independent of the spatial coordinate $h$. We note that live bacteria are swimming at speeds as large as $10-20~\mu$m/s, which is two orders of magnitude larger than the speed of the sedimentation front ($\sim 0.1 \mu$m/s). It is reasonable then to assume that the motion of the live bacteria is unlikely to be affected by the motion of the passive particles or the propagation of the sedimentation front. Solving Eq.~\ref{eq:death_bac} gives $\phi_{b,l}(t)=\phi_{b0}e^{-kt}$, which combined with Eq. \ref{eq:diffu_eqn} leads to $D_{eff}(t)=D_0+ D_1 \phi_{b0}e^{-kt}$ or $(D_{eff}(t)-D_0)=\ln (D_a)=\ln (D_1\phi_{b0})-kt$; here $\phi_{b0}$ is the initial concentration of the swimming bacteria. The quantity $k$ or bacteria loss of activity rate can now be obtained by fitting the above expression to our experimental data in Fig. 3(a). The best fit to the data yields $k=6\times 10^{-6}/$s. This value indicates that at least some bacteria will be active for over 24 hours. While \textit{E. coli} can survive for many hours in different media \cite{ballantyne_1930, schwarz-linek_escherichia_2016}, our MSD data shows that some of it can survive for over a day without additional nutrients. Our measurements are in the range of reported values in the literature for \textit{E. coli} and other gram-negative bacteria species in salt solutions \cite{vaccaro_viability_nodate-1,wijnker_antimicrobial_2006}.

Next, we investigate the effects of activity on (particle) sedimentation front. Figure \ref{fig_vel}(a) show the height of the sedimentation front $h$ normalized by the front initial position $h_{0}$ as a function of sedimentation time for $\xi=\phi_{b0}/\phi_{p0}$ ranging from 5.7 to 22.9. Results show that, for all cases, the sedimentation front decreases linearly as a function of time (at least initially) and slows down significantly as bacteria concentration ($\phi_{b0}$) is increased; the sedimentation front will slow down and develop an exponential form at long times.  As we will show below, the sedimentation front data can also be used to obtain the bacterial motility loss rate $k$ and establish a form of the hindering settling function $H(\phi)$ for active suspensions. 

A relationship between $h$ and bacterial motility loss rate $k$ can be obtained by assuming again Batchelor's settling function \cite{batchelor_1972} for the particle sedimentation speed such that $v_p=v_s(1-n\phi_{b,l})$, which using  Eq.~\ref{eq:death_bac} leads to $v_p(t)=v_s(1-n\phi_{b0}e^{-kt})$. The quantity $h$ can then be expressed as 
\begin{equation}
	h(t)=\int_{0}^{t}v_p~dt=h_0-v_s(t-n\phi_{b0}\frac{1-e^{-kt}}{k}).
	\label{eq:ht_eqn}
\end{equation}

There are two unknowns in the above equation, namely $v_s$ and $n$ (for $\xi > 1$) . These quantities can be obtained by measuring the particle sedimentation speed $v_p$ as a function of (initial) bacterial concentration $\phi_{b0}$ at short times, i.e. $kt\ll 1$. Figure \ref{fig_vel}(b) shows that the sedimentation front speed $v_p$ decreases nearly linearly as the the concentration of live bacteria $\phi_{b0}$ increases. The data shown in Fig. \ref{fig_vel}(b) could be described by an expression of the form $v_p(\phi_{b0})= v_s(1-n\phi_{b0})$; this expression is shown by the solid line in the figure where $v_s =v_p(\phi_{b0}=0)\approx 0.08~\mu m/s$ and $n\approx 120$. This linear dependence is reminiscent of Batchelor's hindering settling function except that we find an unusually large value of $n$. This suggests a dramatic arrest in the particle sedimentation dynamics in the presence of swimming bacteria, likely due to long-range hydrodynamic interactions produce by swimming bacteria. For comparison, Batchelor's original formulation  found $n$ to be equal to 6.5 for passive particles (first order in particle concentration). A slightly better fit to the data is found with an equation of the type $v_p(\phi_{b0})= v_s(1-n\phi_{b0}+((n/2)\phi_{b0})^2)$ with $n=120$, which suggests that second order effects may not be significant. Overall, these results suggest a form of the hindering settling function for active suspensions as a function of bacteria concentration for $\xi>1$ cases. The large value of the constant $n$ for active fluids, $\sim O(100)$, compared to the purely passive case, $n=6.5$  \cite{batchelor_1972,brzinski2018observation}, highlights the role of activity in hindering the sedimentation of particle suspensions.

Since the value of $n$ and $v_s$ are now known, we can proceed to use Eq. \ref{eq:ht_eqn} and the data shown in Fig. \ref{fig_vel}(a) to obtain the quantity $k$, for each $\phi_{b0}$ case. We find that the best fit to our data (all cases presented in Fig. 4a) yields $k=1.0\times10^{-6}/s$ (lines in Fig. \ref{fig_vel}b), which is in the same order of magnitude of the value obtained by measuring the mean square displacement ($k=6\times 10^{-6}$/s). This analysis seems to corroborate the idea that bacteria are dying or losing motility with sedimentation time.  We note that increasing $k$ by 10 times does not have much effect on the profiles of $\bar{h}(t)$, suggesting that our estimate of $k$ from two different methods has the correct order of magnitude. Henceforth, we use $k=6\times 10^{-6}$/s. 

\subsection{Modeling Active Sedimentation}
We now propose a model to describe the concentration profiles measured during sedimentation for $\xi>1$ cases using a modified advection-diffusion equation.  The model is based on two main assumptions. The first is that live bacteria in the suspension have a finite life span due to finite amount of nutrients (and oxygen) and that their loss of activity is a first order process (see Eq.\ref{eq:death_bac}); dead bacteria behave like passive particles. Second, the concentration of live bacteria ($\phi_{b,l}$) is constant throughout the height of the vial $h$, and they die at a constant rate independent of depth and time. 

\begin{table}[h]
\small
  \caption{Some symbols used in sec.B}
  \label{table:symbols_large_disp}
  \begin{tabular*}{0.48\textwidth}{@{\extracolsep{\fill}}lll}
    \hline
   Symbol & Description  \\
    \hline
$h$ & Coordinate along the height of the vial\\
		$h(t)$ & Height of the sedimentation front of passive particles\\
		$\phi_{b,l}$ & Concentration of live \ecoli \\
		$\phi_{b,d}$ & Concentration of dead \ecoli \\
		$\phi_p$ & Concentration of passive particles\\
		$\phi_{b0}$ &Concentration of live \ecoli at $t=0$\\
		$\mathcal{D}_p$ & Dispersivity of the passive particle front\\
		$D_{eff}$ & Diffusion coefficient\\
		$k$ & bacteria loss of motility rate\\
		$\langle r^2(t)\rangle$ & Mean square displacement of passive particles\\
		$v_p$ & $=v_s(1-p\phi_{b,l})$ Speed of the sedimentation front of passive \\& particles\\
		$L$ & Height of the vial\\ 
    \hline
  \end{tabular*}
\end{table}

There are three species in the suspension each of which follows different transport dynamics. They are (i) live bacteria $\phi_{b,l}$, (ii) dead bacteria $\phi_{b,d}$, and (iii) passive particles $\phi_p$. These can be classified into non-active ($\phi_{b,d}$ and $\phi_{p}$) and active ($\phi_{b,l}$) species. The sedimentation process is modelled using a modified version of the advection-diffusion equation (see Eq. 1) that accounts for bacteria loss of activity during sedimentation. In what follows, we describe the dynamics of each specie.  
 
 \textit{Active species} $(\phi_{b,l}(t))$:  The time varying (i.e. decaying) population of live bacteria $\phi_{b,l}(t)$ is described using a first order differential equation (Eq. \ref{eq:death_bac}) that leads to $\phi_{b,l}(t) = \phi_{b0}\exp(-kt)$, where $\phi_{b0} = \phi_{b,l}(t=0)$ is the concentration of live bacteria at time $t = 0$ and $k=6\times 10^{-6}$/s is the bacteria motility loss rate measured using the MSD data. Here, we assume that living \textit{E. coli} are distributed uniformly throughout the bottle, since they are actively swimming at speeds ($10-20~\mu$m/s). This speed is at least two orders of magnitude larger than the typical magnitudes of terminal speeds of the passive particles ($\sim 0.1 \mu$m/s). 
	
    \textit{Passive Species} [$ \phi_{b,d}, \phi_{p}]$: Here, we describe the concentration dynamics of dead bacteria and passive particles during the sedimentation process.  In our experiments, polystyrene spheres represent the passive particles and their transport is governed by a 1-D, time-dependent advection-diffusion equation
\begin{equation}
		\frac{\partial \phi_p}{\partial t} + \frac{\partial( v_p \phi_p)}{\partial h} = \frac{\partial}{\partial h}\left(\mathcal{D}_p\frac{\partial \phi_p }{\partial h}\right),\\
		\label{conv_diff_pass}
	\end{equation}where $\mathcal{D}_p$ and $v_p$ are passive particle dispersivity and sedimentation front speed, respectively. A no-flux boundary condition is imposed at the bottom of the bottle $h=0$ such that
\begin{equation}
		\mathcal{D}_p\frac{\partial \phi_p}{\partial h} -v_p \phi_p=0,
	\end{equation} while the condition $h=L$, $\phi_P(h=L,t)=0$ is enforced at the top of the bottle. 
	
Dead bacteria are assumed to behave like passive particles. These new passive particles (dead bacteria) are constantly being created at all $h$ and $t$. This behavior can be captured by a source term, $\phi_{b,l}(t) = \phi_{b0}\exp(-kt)$, on the right hand side of the advection-diffusion Eq. \ref{conv_diff} which leads to the following expression for the concentration of dead bacteria:
	\begin{equation}
		\frac{\partial \phi_{b,d}}{\partial t} + \frac{\partial(v_b \phi_{b,d})}{\partial h} = \frac{\partial}{\partial h}(\mathcal{D}_b\frac{\partial \phi_{b,d}}{\partial h}) + k\phi_{b0}\exp(-kt).
		\label{conv_diff_bact}
	\end{equation}
Here, $\mathcal{D}_b$ is the dispersivity and $v_b$ is the sedimentation front speed of the dead bacteria. The solution of the partial differential equation above requires two boundary conditions and an initial condition. A no-flux boundary condition is imposed at the bottom of the bottle $h = 0$ such that:
	\begin{equation}
		\mathcal{D}_b\frac{\partial \phi_{b,d}}{\partial h} -v_b \phi_{b,d}=0.	
	\end{equation}
 At the top of the bottle we enforce the condition $\phi_{b,d}(h=L,t) = 0$. At $t=0$, all the bacteria are alive, hence the initial condition is $\phi_{b,d}(h,t=0) = 0$. 
 
 The speeds $v_b$ and $v_p$ in the transport equations given above (Eqs. \ref{conv_diff_bact} and \ref{conv_diff_pass}) depend on the concentration of active bacteria $\phi_{b,l}$. We ignore the effects of passive particle concentration on $v_p$ and $v_b$ because of two reasons: the concentrations of passive particles is constant ($=0.04\%$) in all vials and the concentration is very dilute. Thus, we assume $v_p=v_s(1-n\phi_{b0})$, as shown in Fig. \ref{fig_vel}. We assume the same form for the sedimentation of dead bacteria, \textit{i.e.} $v_b=v_{sb}(1-n\phi_{b0})$. We tested this assumption by performing experiments with UV-immobilized bacteria (not shown) and found that sedimentation speed of passive particles was not significantly affected by dead bacteria; $v_p$ was approximately 15\% smaller for $\phi_{b,d}=0.5\%$. We note that most of our experiments run for 48 hours, which means that about 60\% of our bacteria would be “inactive” by the end of the run. Therefore, we believe that this assumption is reasonable.

While the shape of \textit{E. coli} is rod-like with length $1~\mu m$ and diameter $2~\mu m$, and thus experiences an anisotropic drag, here we will  we assume \textit{E. coli} to be spheres with effective diameter of $d^b=1.44~\mu m$ for the sake of simplicity. The difference in density for \textit{E. coli} and surrounding solution $\Delta \rho^b$ is assumed to be similar to the difference in density for polystyrene and the solution $\Delta \rho^p$, and the terminal speed of a bacterium is then proportional to the square of the effective diameter. Thus, we obtain $\frac{v_{sb}}{v_s}=(\frac{d^p}{d^b})^2\approx \frac{1}{2}$ which implies $v_{sb}=0.06\mu m/s$. We verify the result by manually tracking the dead bacteria front. We find that $v_{sb}=0.055~\mu$m/s, which is in the right range. Even if we double or half this value, the results from our model do not significantly change. Since $v_{b}(\phi_{b,l}(t))=v_{b}(t)$ is a function of time $t$ only, $\frac{\partial (\phi_{b,d} v_{b})}{\partial h}=v_b\frac{\partial \phi_{b,d}}{\partial h}$ in Eq.(\ref{conv_diff_bact}). Similarly, we treat $v_p$ to be devoid of (appreciable) spatial gradients and $\frac{\partial (\phi_p v_{p})}{\partial h}=v_p\frac{\partial \phi_p}{\partial h}$ in Eq  (\ref{conv_diff_pass}). Finally, for simplicity, we assume that dispersivities $\mathcal{D}_p=\mathcal{D}_b$. Here, we note that changing the dispersivities by some amount ($\sim 10\%$) does not have any noticeable effect on the concentration profiles. 

We now integrate the partial differential equations (Eqs. \ref{conv_diff_bact} and \ref{conv_diff_pass}) along with the associated boundary conditions to obtain $\phi_{b,d}(h,t)$ and $\phi_p(h,t)$. Fig. \ref{Fig_prof_low}(c,d) shows that the model is able to capture the main features of the experimental data reasonably well. Our analysis indicates that population dynamics, i.e. accounting for changes in activity, is an important feature in describing the sedimentation of fluids containing living organisms, particularly at long times and for relatively high concentrations. 
\begin{figure}[]
\centering
  \includegraphics[width=.45\textwidth]{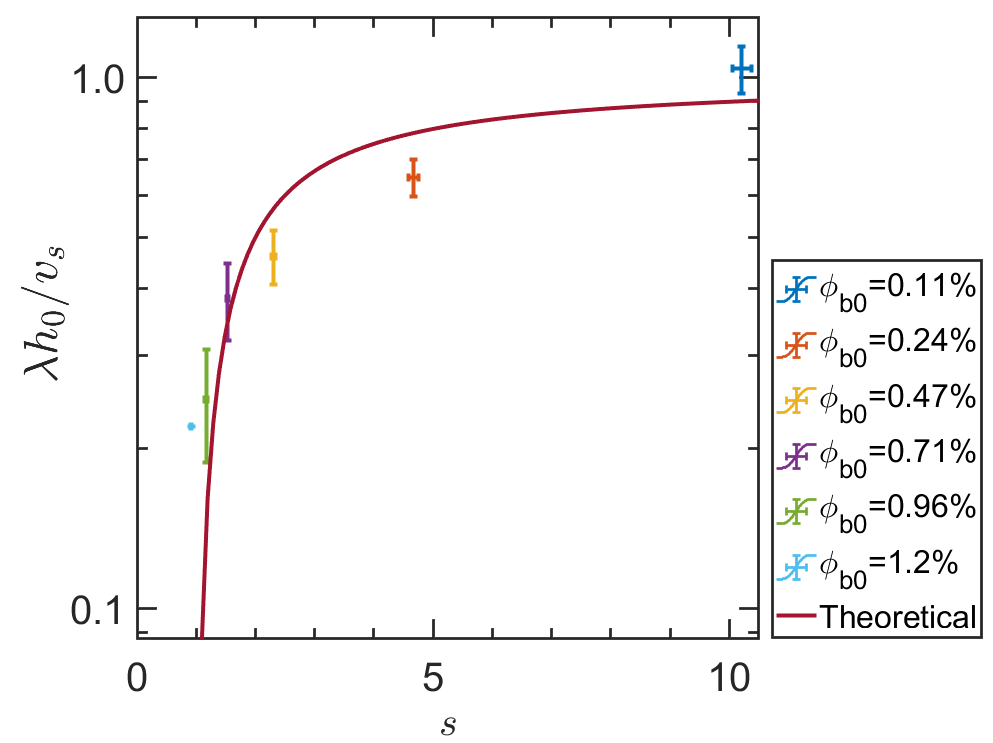}
  \caption{Scaling show the relationship between normalized particle sedimentation speed characterized by $\frac{\lambda h_0}{v_s}$ and bacterial upward flow speed characterized by the quantity $s$ for different values of $\phi_{b0}$. Here, $s = \frac{v_s}{v_s-v_p}$ quantifies the upward flow caused by live bacteria in the suspension, and $\lambda = \frac{\partial N/N_0}{\partial t}$ is the change in the fraction of passive particles suspended in the solution. Our analysis indicates that $\frac{\lambda h_0}{v_s} = 1 - \frac{1}{s}$. }
  \label{scaling_fig}
\end{figure}

\subsection{A Simple Scaling}
The strong hindering of (passive) particle sedimentation in the presence of swimming microorganisms in our experiments may be due to passive particles experiencing a bias in vertical (upward) velocity fluctuations produced by the swimming \textit{E. coli}; bacteria may be preferentially moving towards the oxygen-rich portion of the bottle (i.e. top). This bacteria flow, even in the dilute regime, could to be enough to keep particles re-suspended in the fluid for longer periods of time compared to the case of no bacteria. Fluid flows are known to keep particles re-suspended in liquid media (e.g. fluidized beds and mixing tanks). For example, the settling of crystals in a convecting magma chamber is found to be hindered by a random flow due to cooling from above \cite{martin1988cryst}; convective velocities greatly exceed the settling speeds throughout most of the depth of the chambers away from the walls. Similarly, bacteria swimming speeds are typically much larger than particle sedimentation speeds, and we hypothesise that swimming bacteria in the vials may create flows with velocity fluctuations that are vertically biased that may keep particles suspended in the fluid leading to the dramatic arrest in particle sedimentation observed in our experiments. 

Here, we describe the sedimentation process in the presence of live bacteria using two non-dimensional (speed) parameters $s$ and $\frac{\lambda  h_0}{v_s}$, where $h_0$ is the initial sedimentation height and $v_s$ is the sedimentation speed of a single (passive) particle; this is analogous to \cite{martin1988cryst}. The quantity $s$ is a non-dimensional characteristic speed that quantifies the flow caused by the presence of live bacteria, $s= \frac{v_s}{v_s-v_p}$; we use the slow down in sedimentation front speed $v_s - v_p$ as an estimate of the flow produced by bacteria. Substituting $v_p = v_s(1-n\phi_{b,l})$ and noting that $\phi_{b,l}(t) = \phi_{b0}\exp(-kt)$ leads to $s = \frac{1}{n\phi_{b0}e^{-kt}}$, where $n$ $\approx$ 120 and $k=6\times 10^{-6}$/s is the bacteria motility loss rate. The quantity $\lambda$ is a characteristic time-scale that describes the decay in the fraction of passive particles ($N/N_0$) in the solution at time $t$, which for our experiments is given by $\lambda = \frac{\partial N/N_0}{\partial t} = \frac{1}{h_0}\frac{\partial \bar{h}(t)}{\partial t}$, where $N \propto h$ while $ N_0 \propto h_0$. Using Eq. \ref{eq:ht_eqn} gives, $\frac{\lambda h_0}{v_s} = 1-\frac{1}{s}$. This implies that our data for $ \frac{\lambda h_0}{v_s}$ vs $s$, should collapse onto the curve $y = 1 - \frac{1}{x}$ for different values of $\phi_{b0}$. 

Figure \ref{scaling_fig} show experimental data for $\phi_{b0}$ ranging from 0.012 $\%$ to 0.94 $\%$ (and $\xi$ from 0.28 to 22.9). The solid line in Fig. \ref{scaling_fig} shows that the scaling seems to capture our data relatively well, thus providing support for the assumptions in Eqs. \ref{conv_diff_bact} and \ref{conv_diff_pass}. Moreover, this analysis suggests that (i) there may be an upward flow created by swimming bacteria that keeps particle suspended in the fluid even in the dilute regime and (ii) the sedimentation process can be captured by the ratio of the bacterial flow to the particle Stokes' settling speed, provided that population dynamics or changes in activity are taken into account. 
\begin{figure}
 \centering
 \includegraphics[height=4.2cm]{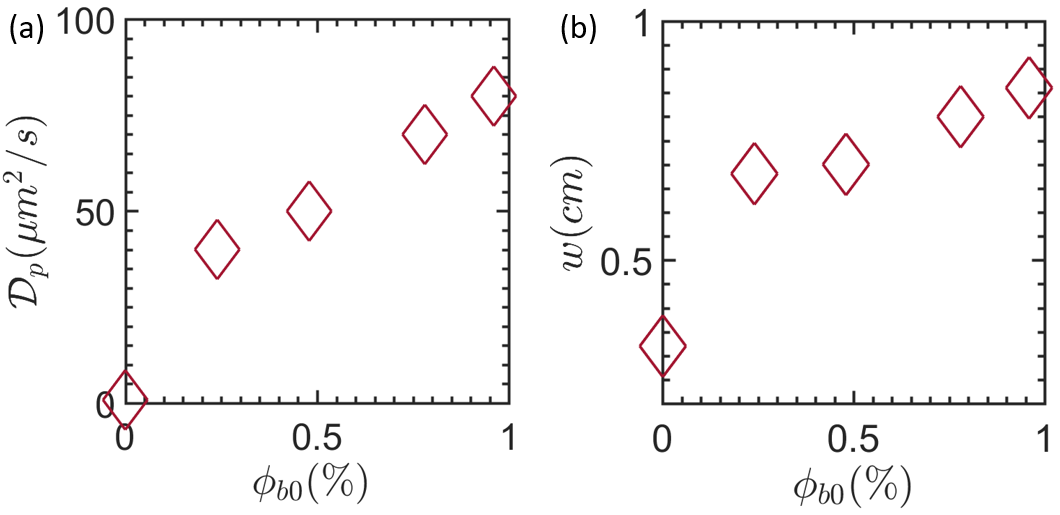}
 \caption{(a) Dispersivities $\mathcal{D}_p$ as a function of bacterial concentration (b) The width of the sedimentation front at $t=25$ hr as a function of the bacteria concentration. Fitting error is approximately 5\%.}
 \label{Fig_diff}
\end{figure}

In summary, we find that the effect of the presence of live bacteria in settling particle suspensions is two-fold: (i) the speed of the sedimentation front decreases with increasing concentration of live bacteria, and (ii) the dispersivity, in case of suspensions containing large concentration of live bacteria (figure \ref{Fig_prof_low}), are much larger ($\sim10$ times) than those observed in cases where bacteria are either absent or present in small concentrations (Fig \ref{Fig_prof_low}). The variation of the fitted dispersivities with live bacteria concentration is shown in figure \ref{Fig_diff}a. The dispersivity increases with the concentration of live bacteria, consistent with the corresponding increase in the width of the sedimentation front (figure \ref{Fig_diff}b).

\section{Conclusions}
The sedimentation of passive particles in the presence of live bacteria is investigated both in experiments and using a simple model. We find that the presence of swimming bacteria significantly hinders the sedimentation of passive particles. Even at low concentrations of live bacteria ($\phi_{b} = 0.012 ~\%$ ), we find that the presence of bacteria increases the dispersivity of the passive particles, while the mean sedimentation speed remains unchanged. As the concentration of bacteria $\phi_{b}$ increases, we observe strong deviations from this behavior: the dispersion coefficient of the passive particles increases with $\phi_{b}$ (Fig \ref{Fig_diff}a) and the sedimentation speed decreases rapidly compared to passive particle suspensions, even for concentrations of particles and bacteria considered dilute ($\phi < 1\%$) (figure \ref{fig_vel}a). Moreover, we find a decrease in live bacteria population (or activity) with sedimentation time.  Our model suggests that a source term representing this population change over time needs to be included in order to capture the experimental data. That is, an advection-diffusion systems of equations with a source term yields a reasonable model for sedimentation of active suspensions. The key ingredients are that (a) the particle speed on the left hand side of Eq. \ref{conv_diff} is a function of live bacteria concentration that also varies with time, and (b) a time dependent source of passive particles also appears in the governing equation due to bacteria loss of activity/motility. We find that, at least in the dilute regime, our experimental sedimentation data is captured by the ratio of bacterial (upward) flow in the vial to the sedimentation speed of a single passive particle. The scaling includes the decay of live bacteria over time. 

Our study has implications for describing the sedimentation process in which active particles are present. We have shown that, in describing such active systems, population dynamics of bacteria cannot be ignored. Here, we have treated the population dynamics of the isolated bacteria in a simple manner and shown that it was sufficient to account for the observations in experiment. However, more sophisticated treatments might be necessary to account for a motility loss rate $k$ that is time and spatially dependent and when the bacteria are not isolated as in our vials. More broadly, our study could have implications on sedimentation processes in geological and man-made water reservoirs in which live micro-organisms are almost always present. A natural next step would be to explore the role of the particle size in sedimentation, since larger particles can diffuse faster than smaller particles in suspensions of swimming bacteria \cite{Patteson2016}; this effect could lead to anomalous sedimentation speeds and diffusion coefficients, which may control particle sorting during sedimentation. 

\section*{Conflicts of interest}
There are no conflicts to declare.

\section*{Acknowledgements}
We thank A. Gopinath, J. Bush, D. Durian, S. Spagnolie, A. Baskaran, and R. Radhakrishnan for fruitful discussions. A. E. Patteson, B. Maldonado, and P. E. Arratia acknowledge support by the National Science Foundation grant DMR-1709763. J. Singh and P. K. Purohit acknowledge support through National Science Foundation grant DMR-1505662 and the University Research Foundation at the University of Pennsylvania. A. E. Patteson was supported by NSF Graduate Research Fellowship.\\

\bibliography{sedimentation} 
\bibliographystyle{rsc}

\end{document}